\documentstyle[prb,aps,psfig]{revtex}

\begin{document}
\draft

\title{A theory of ferromagnetism in  planar heterostructures of
(Mn,III)-V  semiconductors}
\author{J. Fern\'andez-Rossier and  L. J. Sham }
\address{Department of Physics, University of California San Diego,
 9500 Gilman Drive, La Jolla, CA 92093-0319}

\date{\today}

\twocolumn[\hsize\textwidth\columnwidth\hsize\csname@twocolumnfalse\endcsname
\maketitle

\begin{abstract}

A density functional theory of ferromagnetism in  heterostructures of
compound  semiconductors doped with magnetic impurities is
presented.  The variable functions in the density  functional theory
are  the charge and spin densities of the itinerant  carriers and  
the charge and localized spins of the impurities. The theory is
applied  to study the Curie temperature of planar heterostructures of
III-V semiconductors doped with manganese atoms. The mean-field, 
virtual-crystal and effective-mass approximations are adopted to
calculate the  electronic structure, including the spin-orbit
interaction, and the magnetic susceptibilities, leading to the Curie
temperature. By means of these results, we attempt to understand the
observed dependence of the Curie temperature of planar $\delta$-doped
ferromagnetic structures on variation of their properties. We predict
a large increase of the Curie Temperature by
additional  confinement of the holes in a $\delta$-doped layer of Mn
by a quantum well.

\end{abstract}


]

\narrowtext

\section{Introduction}

Interest in ferromagnetic III-V semiconductors lies both in
fundamental physics and potentially useful technological applications
utilizing spins. \cite{Spintronics}  The  growth   of
Ga$_{1-x}$Mn$_{x}$As, a  ferromagnetic III-V semiconductor,
\cite{Ohno96,Esch97,Ohno98a,Ohno98b,Ohno00} has raised the basic
problems of the origin of the ferromagnetism and  of the spin
transport properties. In contrast to the much studied Mn-doped II-VI
materials, \cite{Furd} Mn acts as an acceptor in GaAs so that
Ga$_{1-x}$Mn$_{x}$As has free holes which are thought to be
responsible for the high Curie temperature \cite{Ohno96}  ($T_C$) of 
110~K for $x=0.054$.   Progress in ``spintronics'' is made by the
recent demonstrations of the injection of a spin-polarized current
from both ferromagnetic metals \cite{spinjec1,spinjec2,spinjec3,spinjec4} 
and magnetic semiconductors \cite{spinjec5,spinjec6}
into a semiconductor.

The Zener model of ferromagnetism for a bulk alloy of (III-V)
semiconductors, \cite{Dietl97,Dietl00} linking the Curie Temperature
with the spin susceptibility of the mobile holes, forms a framework
for understanding the dependence of the ferromagnetism on various
properties of the system, including the hole density and the Mn
concentration. The semiconductors afford an opportunity to change the
properties to affect the ferromagnetism, for example, by optical
excitation of the carriers \cite{koshi} and by  field effect control.
\cite{Naturedec2000}  Engineering  the band gap  profiles of planar
heterostructures of the III-V semiconductors and the doping can also
vary the factors influencing the  ferromagnetic order. In 
experiments \cite{Akiba} in   a double quantum well of GaAsMn/GaAsAl,
the coupling between the two magnetic layers is observed to depend on
both the thickness and the composition of the non-magnetic barrier.

A different kind of planar system, the so-called digital
ferromagnetic heterostructure, has been introduced by R. Kawakami
{\em et al.} \cite{Kawakami} This system consists of a sequence of
atomic monolayers of   Ga$_{1-x}$Mn$_{x}$As (with x=0.25 and 0.5)
separated by several layers of GaAs. This  $\delta$-doping
\cite{Deltabook} structure displays a transition temperature which is
a decreasing function of the distance between the magnetic layers. At
the shortest distance reported, $T_C$ is around  50~K.
\cite{Kawakami}  At large interlayer distance, the layers are
decoupled and $T_C$ reaches a value of around 35~K. Single layer
samples also show ferromagnetism. \cite{private}

In this paper we provide a theoretical framework to study
ferromagnetism in planar heterostructures of ferromagnetic III-V
semiconductors. The model combines the standard procedure to
calculate the electronic structure of planar semiconductor
heterostructures in the envelope function formalism with a mean field
theory for the ferromagnetic state. Our goal is to understand the
interplay of confinement, spin-orbit interaction, and close packing
of Mn atoms in the ferromagnetic  digital heterostructures. Keeping
the theories of the electronic structure and  of ferromagnetism
simple enable us to study the effects of varying the  system
configurations, degree of interdiffusion, and carrier compensation. 
We  apply the theory to calculations of the electronic structure and
$T_C$ for the case of single and multiple digital layers of GaAsMn
and  for the additional confinement effect by a quantum well of a Mn
layer. We  show that the behavior of the double layer as a function
of the layer  separation is essentially that of more than two layers.

Our model is an extension  to the case of planar heterostructures of
the Zener model for bulk alloys of (III-V) ferromagnetic
semiconductors. \cite{Dietl97,Dietl00} It is convenient  to express
the problem of the localized spins and the carrier spins  in an
inhomogeneous systems in terms of the density functional theory. 
\cite{KS} It expresses the Curie Temperature in terms of the carrier
spin  susceptibility of the doped semiconductor, taking into account
the spin-orbit  interaction of the holes. Such an extension involves
two technical refinements,  compared to the bulk case. First, as
confinement breaks the translational  invariance along the growth
axis $z$,  the susceptibility becomes a non-local  function
$\chi(z,z')$. In order to  determine $T_C$ in a planar
heterostructure  we have to solve an integral equation whose kernel
contains the non-local  spin susceptibility. The derivation of the
integral equation from the  density functional theory is given in
section II.  Second, as explained in  section III, the calculation of
the electronic structure of the holes in a  planar heterostructure
involves the  solution of the self-consistent Schr\"odinger equation.
\cite{Broido} The calculation of the subbands, for the in-plane
motion, is made using the ${\bf k}\cdot\vec{p}$ Luttinger
Hamiltonian, \cite{KohnLutt} which includes the spin-orbit
interaction. Our calculations address the  regime in which both the
Mn density, $c_{M}$, and the  carrier density, $p$, are high.  This
implies that several subbands are occupied. Therefore, our
calculations include {\em both} spin-orbit interaction  and multiple
sub-bands, in contrast with previous work. \cite{Mac00,MacQW,brey,nosoqw}

In section IV we apply the formalism to the case of a single
digital layer of GaAsMn embedded in GaAs. From the calculated
electronic structure for different values of the degree
of compensation and the interdiffusion of the Mn,  we find that
the calculated $T_C$ is an increasing function the density of holes and 
a decreasing function of the interdiffusion of the magnetic
impurities. In section V, we investigate the change in $T_C$  as a 
function of the interlayer distance between two layers in order to understand 
the observed behavior for multilayers of Mn for different degrees of 
compensation and interdiffusion. Our numerical results for two layers,
which are very similar to those for up to five
layers, reproduce the main features of the experimental results. 
A remarkable  result is found in section VI where the calculated $T_C$
for a digital layer  inside a quantum well is found to increase
dramatically by the additional  quantum well confinement, compared to
just the confinement due to the Mn layer in  section IV. In section
VII we put our results in perspective and draw some conclusions.

\section{Density functional formulation for magnetism in 
heterostructures}

 The observed effects of ferromagnetism in Mn-doped III-V 
semiconductors appear to be consistent with the microscopic mechanism
of indirect  Mn-Mn spin interaction mediated by the mobile holes via
the exchange  interaction between the hole and the magnetic moment of
the localized d electrons  of the Mn impurity. A common model for the
bulk system, which we shall adopt, consists of a quantum degenerate  
gas of fermions  which interact, via  a contact Heisenberg exchange
interaction, with the local magnetic  moments of Mn. The random array
of Mn impurities is replaced  by a homogeneous distribution and the
exchange interaction is  reduced correspondingly. This procedure is
dubbed the  virtual crystal approximation. \cite{Furd}  The exchange
coupling between Mn and the holes is treated in the mean-field
approximation. \cite{Dietl00}

The study of heterostructures of semiconductors $\delta$-doped with
manganese, calls for an  extension of  the virtual  crystal and
mean-field approximations to an inhomogeneous distribution of Mn.
We formulate the theory of magnetism of the inhomogeneous system in 
terms of the density functional theory, extended to include the spin
densities  and to finite temperature. \cite{KS} The free energy of
the system of Mn spins and holes as a functional of the density and
spin density distributions of the Mn spins and the hole carriers is
separated into  three contributions:
\begin{equation}
F  = F_M +E_h +E_{hM},
\end{equation}
respectively, of Mn spins, the holes, and the Mn-hole interaction.
The free energy of the Mn spin system alone is:
\begin{eqnarray}
&&F_M[c_M({\bf r}), {\bf M}({\bf r})] \nonumber \\
&=& \int d^3r \, c_M({\bf r}) [ f_0(M({\bf r}))
+ \case1/2 J_M M^2({\bf r})] \nonumber \\
&+& \case1/2\int d^3r\int d^3r' \, c_M({\bf r})c_M({\bf r}')
J_{MM}({\bf r}, {\bf r}') M({\bf r}) M({\bf r}')] .
\end{eqnarray}
The number density of the Mn is denoted by $c_M({\bf r})$ and ${\bf 
M}({\bf r})$ is the spin expectation per Mn atom (in units of
$\hbar$). The noninteracting part of the free energy per Mn spin is
given by
\begin{equation}
f_0({\bf M}) = k_BT \left[Mb - \ln
\left\{\frac{\sinh(S+\case1/2)b}{\sinh\frac{b}{2}} \right\} \right],
\end{equation}
where $k_B$ is the Boltzmann constant, $S=\case5/2$ is the spin of
Mn  and $b=b(M)$ is related to the inverse of the usual Brillouin
function,
\begin{equation}
M =  (S+\case1/2) \coth[(S+\case1/2)b] - \case1/2
\coth\left(\case1/2 b\right).
\end{equation}
The second term in the integral takes into account only the
short-range Heisenberg exchange between Mn spins. $J_M$ has the
dimension of  energy. In GaAs it is believed to be antiferromagnetic.
\cite{Dietl00} The  third term accounts for the long-range dipole
interaction $J_{MM}({\bf r}, {\bf r}')$ and is found to be 
negligible.

The temperature range under study is sufficiently low compared with
the Fermi temperature of the holes that the hole free  energy  will
be taken as the usual ground state energy functional $E_h[p({\bf r}),
{\bf S}({\bf r})]$, \cite{KS} where $p({\bf r})$ is  the hole density
and ${\bf S}({\bf r})$ the spin density.  The interaction between 
holes is included in the Hartree approximation in the  calculation of
the sub-bands. As the density of holes is very high, the exchange 
and correlation potential in the local-density approximation is of
minor importance.

The Mn-hole interaction term is given by
\begin{eqnarray}
&& E_{hM}[p({\bf r}), {\bf S}({\bf r}); c_M({\bf r}), {\bf M}({\bf r})]
\nonumber \\ &=& - \int d^3r \int d^3r' \, p({\bf r}) u({\bf r}-{\bf 
r}')
[c_M({\bf r'}) - c_c({\bf r'})] \nonumber \\
 &+&J \int d^3 r\; c_M({\bf r}) {\bf M}({\bf r}) \cdot {\bf S}({\bf r}).
\end{eqnarray}
The first term on the right side of the equation is the attractive 
potential provided by the Mn donors to the holes with $u({\bf r}-{\bf
r}')$ being  the Coulomb interaction. Experiment \cite{Ohno96} shows
compensation of the acceptors, which are believed to be antisite
impurities. The compensating impurity concentration $c_c({\bf r})$ is
taken into account. The second  term is the hole spin interaction
with the Mn spin, for which we use the  simplest mean-field
approximation. Functional terms beyond the mean field might  be
constructed from theory such as Ref.~\onlinecite{Mac00b,Yangcomm,KL}. The
hole-Mn  spin interaction $J$ has the dimension of energy-volume.

In the mean-field and virtual-crystal approximations, the model
describes the hole carriers interacting with an effective magnetic
field produced by the localized Mn impurities and vice versa. The 
variational result of the free energy functional with respect to both
the  magnetization of the Mn impurities and the magnetization of the
hole carriers shows  the interdependence of the two magnetizations.
Each is governed by the  effective magnetic field generated by the
other. They have to be determined selfconsistently. We report here
only the work on the transition temperature.  Theoretical finite
magnetization studies are being carried  out. Close to the  Curie
temperature, the magnetizations are small. The free energy is  a 
quadratic functional of the two magnetizations. In a planar
heterostructure with the growth axis along $z$, there is
translational invariance along the $x-y$ plane in the effective mass
approximation so that quantities depend only on $z$.  The free energy
functional per unit area is then:
\begin{eqnarray}
&F&[p(z), {\bf S}(z); c_M(z), {\bf M}(z)]
= F[p(z),c_{M}(z)]+ \nonumber \\  &+&  \case1/2 \int dz \;
 c_{M}(z)\left[
 \frac{3 k_B T }{S (S+1)} +J_M \right]
M_{\alpha}(z)^2
 \nonumber \\
&+&J \int \; dz c_{M}(z) M_{\alpha}(z) S_{\alpha}(z)
\nonumber \\
&+& \case1/2 \int dz \; \int dz' S_{\alpha}(z)
K_{\alpha}(z,z') S_{\alpha}(z') .
\label{DFT}
\end{eqnarray}
For simplicity, we investigate easy magnetization of $M_{\alpha}(z)$ 
and $S_{\alpha}(z)$ only along the growth axis $\alpha = z$  or
in plane  $\alpha =x$.  The first term $F[p(z),c_{M}(z)]$ is the
density functional for zero magnetization including the impurity 
potential for the holes. Charge neutrality  determines the total
number of holes:
\begin{equation}
\int_{-\infty}^{\infty} p(z) dz =\int_{-\infty}^{\infty} \left[
c_{M}(z)-c_{c}(z)  \right]dz.
\end{equation}
The first quadratic term in $M$ contains the inverse susceptibility
of  the Mn spins. The last term is the magnetic energy of the holes,
where $K_{\alpha}$ is the inverse of the non-local hole spin
susceptibility $\chi$ \cite{KS}:
\begin{equation}
\int  dy \; K_{\alpha}(z,y) \chi_{\alpha}(y,z')= \delta(z-z').
\label{inverse}
\end{equation}

Minimization of the energy functional (\ref{DFT}) with respect to the
magnetizations leads to two coupled equations for $M_{\alpha}(z)$ and
$S_{\alpha}(z)$:
\begin{eqnarray}
c_{M}(z)\left\{ \left[ \frac{3 k_B T}{S (S+1)} +J_M \right]
+ J S_{\alpha}(z) \right\} &=& 0 , \label{ex1} \\
\int dz' \,
K_{\alpha} (z,z') S_{\alpha}(z') + J c_{M}(z)M_{\alpha}(z) &=& 0.
\label{ex2}
\end{eqnarray}
$M_{\alpha}(z)=S_{\alpha}(z)=0$ are always a solution, corresponding
to  the stable state only in the paramagnetic phase. The Curie
temperature is  the highest temperature at which
Eqs.~(\ref{ex1},\ref{ex2}) have non-zero solutions. Elimination of
$S_{\alpha}(z)$ from  Eqs.~(\ref{ex1},\ref{ex2}) by
Eq.~(\ref{inverse}) leads to an integral equation for the Curie
temperature $T_C$:
\begin{eqnarray}
 &&M_{\alpha}(z)c_{M}(z)= \nonumber\\
&& \frac{c_{M}(z)S(S+1) J^2}{3 (k_B T_C +k_B T_{M})  }\int dz
\chi_{\alpha}(z,z')
   c_{M}(z')M_{\alpha}(z')
\label{integ}
\end{eqnarray}
 where $k_B T_{M}\equiv S(S+1) J_M/3$. Eq.~(\ref{integ}),  which
relates $T_C$  with the non-local spin susceptibility of the holes,
$\chi_{\alpha}(z,z')$, for a given  planar heterostructure is the
main  result of this section. Eq.~(\ref{integ}) extends Eq.~(10--12)
of Ref.~\onlinecite{MacQW} to include multiple subband occupation
and  spin-orbit interaction.

\section{Non-local susceptibility}

The calculation of the Curie Temperature for the planar
heterostructure involves the solution of the integral equation
(\ref{integ}), whose kernel contains the non-local spin
susceptibility. In this section we briefly describe our calculation
of the electronic structure and the non-local spin susceptibility for
a rather general planar heterostructure  characterized by a profile
of the  Mn impurities $c_{M}(z)$, the profile of the compensating
impurities $c_{c}(z)$ (antisites), and a band gap profile which
creates a potential for the holes $V_{i}(z)$.

For a given density profile we solve  the Poisson equation and obtain
the electrostatic potential $V_{el}$, which, together with a band gap
potential $V_{i}(z)$, defines the effective mass Hamiltonian
\cite{Broido,KohnLutt} for the envelope function of the holes:
\begin{equation}
H_{\mbox{eff}}=H_{L}({\bf k},\frac{1}{i}\frac{\partial}{\partial 
z})+V(z)
\end{equation}
 where ${\bf k}$ is the in-plane wave vector,
$V(z)=V_{i}(z)+V_{el}(z)$, $H_{L}$ is the standard 4$\times$4
Luttinger Hamiltonian, with parameters $\gamma_1$,
$\overline{\gamma}\equiv 0.5 (\gamma_2+\gamma_3)$, and $\mu \equiv
0.5(\gamma_3-\gamma_2)$. We adopt  the {\em cylindrical
approximation} about the growth axis \cite{Broido},  i.e., taking
$\mu=0$ in the Luttinger Hamiltonian. This approximation has the
advantage that $\epsilon_{{\bf k},\nu}$ does only depend on $k
\equiv|{\bf k}|$. The physical properties involve angular integration
over ${\bf k}$ about $z$ so that the deviations from the cylindrical
approximations  are very small \cite{yang}.

The subbands $\epsilon_{{\bf k},\nu}$ and the corresponding
eigenstates, $\psi_{{\bf k},\nu}(z)$, are given by the solution of
the set of four coupled second order differential equations:
\begin{eqnarray}
\sum_{m=-\case3/2}^{\case3/2}
\left[H^{\mbox{eff}}_{n,m}({\bf k},\frac{1}{i}\frac{\partial}{\partial 
z}) +
V(z)
\delta_{n,m} \right] F^{{\bf k},\nu}_m(z)=
\epsilon_{{\bf k},\nu}  F^{{\bf k},\nu}_n(z)
\label{lutti}
\end{eqnarray}  where
\begin{equation}
 \left( F^{{\bf k},\nu}_{3/2}(z),
  F^{{\bf k},\nu}_{1/2}(z),
 F^{{\bf k},\nu}_{-1/2}(z),
 F^{{\bf k},\nu}_{-3/2}(z) \right) =\psi_{{\bf k},\nu}(z)
\end{equation}
are the four components of the wave function $\psi_{{\bf k},\nu}(z)$.
Eq.~(\ref{lutti}) is solved with the mini-band ${\bf k} \cdot{\bf
p}$  method \cite{Broido}. We first solve the $|{\bf k}|=0$ case, in
which the  equations are decoupled into two ordinary Schr\"odinger
equations corresponding  to the light and the heavy holes. These
have  $n_h$ and $n_l$ bound states, evaluated by transforming the one
dimensional Schr\"odinger equation  into a tridiagonal matrix
eigenvalue problem which is solved numerically \cite{Lazzouni}. The
$|{\bf k}|=0$ solutions form a basis set with the  $2 n_h$ states
$\psi_{\nu h} (1,0,0,0)$, and $\psi_{\nu h} (0,0,0,1)$ and  the $2
n_l$ states $\psi_{\nu l} (0,1,0,0)$ and $\psi_{\nu l} (0,0,1,0)$. 
The finite $|{\bf k}|$ eigenergies $\epsilon_{{\bf k},\nu}$ and
eigenstates $\psi_{{\bf k},\nu}(z)$ are obtained in terms of the
basis set as the solutions of Eq.~(\ref{lutti}) as the $N\times N$
secular determinant problem, where $N=2(n_h+n_l)$. The hole density
is given by:
\begin{equation} p(z)=\sum_{\nu} \int\frac{d^2k}{(2
\pi)^2}f(\epsilon_{{\bf k},\nu})|\psi_{{\bf k},\nu}(z)|^2
\end{equation} where $f(\epsilon)$ is the Fermi-Dirac occupation
function, the Fermi level being fixed so that the charge
neutrality condition is met.  The hole density $p(z)$ and the potential
$V(z)$ are determined by iteration to self-consistency.

The nonlocal spin susceptibility is then given by:
\begin{eqnarray}
\chi_{\alpha}(z,z') =\sum_{\nu} \int \frac{d^2 k}{ (2 \pi)^2}
\left[ \frac{\partial f(\epsilon_{{\bf k},\nu})}{\partial
\epsilon_{{\bf k},\nu})}  S^{\alpha}_{{\bf k},
\nu,\nu}(z)\; S^{\alpha}_{{\bf k},\nu,\nu}(z')
 \right. \nonumber \\  \left.
 +  \sum_{\nu'\neq\nu} S^{\alpha}_{{\bf k},\nu,\nu'}(z)
 \frac{f(\epsilon_{\nu}({\bf k}))-f(\epsilon_{\nu'}({\bf k}))}
 {\epsilon_{\nu}({\bf k})-\epsilon_{\nu'}({\bf k})}
 S^{\alpha}_{{\bf k},\nu',\nu}(z') \right],
\end{eqnarray} where $\alpha=(x,y,z)$  and the spin matrix
elements are given by:
\begin{equation} S^{\alpha}_{{\bf k},\nu',\nu}(z)= \langle
\psi_{{\bf k},\nu'}(z) |S^{\alpha} |\psi_{{\bf k},\nu}(z) \rangle,
\end{equation} where the angular brackets denote the expectation value 
over the spin degrees of freedom of the hole states $\psi_{{\bf
k},\nu}(z)$.  In the absence of spin-orbit interaction, the spin
matrix elements would be independent of the in-plane momentum ${\bf
k}$. Moreover, due to the interplay between the spin-orbit 
interaction and confinement, the nonlocal spin susceptibility takes
different values  for in-plane and off-plane orientation. If the
spin-orbit interaction was the only source of anisotropy, our
calculation could determine the easy axis. Other sources of
anisotropy, like shape anisotropy, are not considered in our
calculation. In the calculations reported in this paper, we assume an
in-plane magnetization, guided by the experimental result
\cite{Kawakami}.

\section{Single Digital Layer}

In this section we present the results of our calculations of the 
electronic structure and the Curie temperature, $T_C$, for GaAs doped
with a  single digital Mn  layer of (Ga$_{0.5}$Mn$_{0.5}$)As, as in
the experiments. In an ideal $\delta$-doping, the Mn atoms occupy a
single atomic plane. In the real system, Mn atoms undergo
interdiffusion to occupy several layers. We assume that the
compensating impurities are closely associated with the Mn atoms and
assume their distributions to have the  same shape:
\begin{equation} \frac{c_{M}(z)}{\alpha_{M}} =  
\frac{c_{c}(z)}{\alpha_{c}}=\sum_{n}
\frac{1}{\Delta}e^{-(na/\Delta)^2}\delta(z-na),
\label{profile}
\end{equation} where  $\alpha_{c}$ and $\alpha_M$ lead,
respectively, to the total concentrations of Mn, $c_M$ and  of the
compensating impurities, $c_c$ so that the density of holes is
$p=c_{M}-c_c$.   Hence, for a given Mn concentration
$c_{M}$, a single layer is characterized by ($\Delta,p$)

In the limit of $\Delta=0$, we recover the ideal $\delta$-doping
case, $c_{M}(z)=c_0 \delta(z)$.   Then, Eq.~(\ref{integ}) can be
solved analytically: \begin{equation} k_B T_C=\case1/3 c_{M} 
S(S+1)J^2 \chi(0,0)- k_B T_{M} . \end{equation} The dynamics of the
holes is contained in $\chi(0,0)$.  We see that $T_C$ does {\em not}
depend on the sign of the hole Mn exchange interaction, $J$. On the
other hand, $-k_B T_{M}$, proportional to the direct Mn-Mn
interaction, if  antiferromagnetic ($J_M>0$), {\em decreases} the
Curie temperature, as expected. In the  case of a random alloy of
GaAsMn,  $J_M$ is found negligible \cite{Dietl00} because the
distance between the Mn impurities is rather high. In  contrast, the
in-plane average distance between  the Mn is much shorter in the
digital heterostructure. However,  an accurate value for both $J$ and
$J_M$ is not known.  Hereafter, we set \cite{Ohno98b} $J_M=0$ and
$J=150$~meV-nm$^3$.

\begin{figure}
\centerline{\psfig{figure=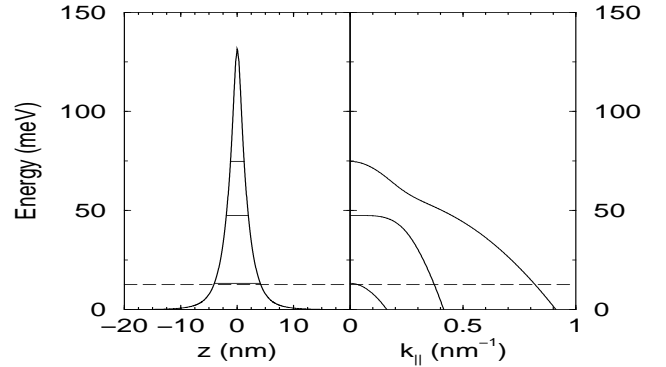,height=2.0in,width=3.3in}}
\caption{Left panel: selfconsistent potential for a single digital
layer with $\Delta=0.5$~nm~ and
$p=1.3 \times 10^{13}$~cm$^{-2}$, including the HH and LH levels.
Right panel: hole subbands. The dashed line is the Fermi level.}
\label{fig1}
\end{figure}

In the left panel of Fig.~\ref{fig1} we plot the selfconsistent
potential for the holes corresponding to a single digital layer with
$\Delta=0.5$~nm~ and $p=1.3 \times 10^{13}$~cm$^{-2}$, together with
the energy levels for the light and the heavy holes.  In the right
panel we plot the subbands for the in-plane motion of the holes. The
dashed line indicates the Fermi level.  For Ga$_{0.5}$Mn$_{0.5}$As, we
have $c_M=3.13\times 10^{14}$~cm$^{-2}$. Even for a density  of holes
at only 4.1\% of the Mn concentration, 3 subbands are occupied
with holes. For this set of parameters the obtained $T_c$ is 35$ K$
the experimental value obtained by R. Kawakami {\em et al.}.
\cite{Kawakami}

The spin-orbit interaction causes both the anticrossing
and the non parabolic shape of the hole subbands. As in  the bulk
case, \cite{Dietl00} the spin-orbit effect also reduces 
significantly the effective magnetic coupling between Mn spins.
Therefore,  it is  important to include spin-orbit in the theory of
ferromagnetism in planar heterostructures, an ingredient missing in
previous papers for quantum wells. \cite{Mac00,MacQW,brey,nosoqw}

\begin{figure}
\centerline{\psfig{figure=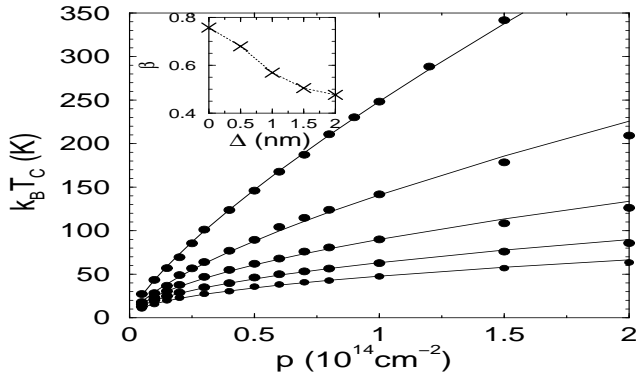,height=2.0in,width=3.3in}}
\caption{ $T_C$ as a function of density of holes,p,  for
$\Delta=$0, 0.5, 1.0, 1.5 and $2.0$~nm~ (from top to bottom). Inset:
$\beta$ as a function of $\Delta$
 for the fit $T_C (p,\Delta) \propto p^{\beta(\Delta)}$. }
\label{fig2}
\end{figure}

\begin{figure}
\centerline{\psfig{figure=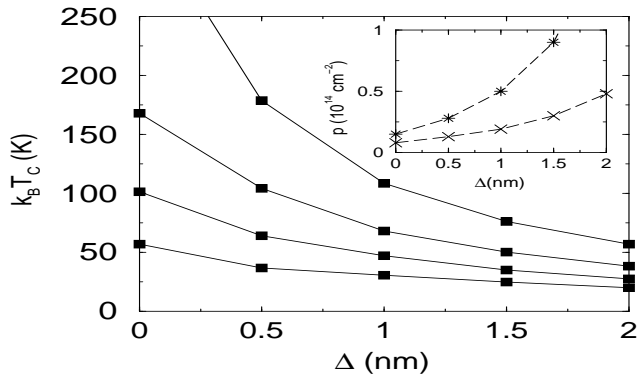,height=2.0in,width=3.3in}}
\caption{ $T_C$ as a function of the interdifusion parameter,
$\Delta$, for $p=1.5$ (highest), $0.6$, $0.3$, $0.15$ (lowest) in
units of $10^{14}$~cm$^{-2}$. Inset:  Lines of $T_C$= 35~K (crosses)
and $T_C=$60~K (stars), in the $(p,\Delta)$ plane (with $J=150$~meV-nm$^3$).}
\label{fig3}
\end{figure}

In Fig.~\ref{fig2} we show $T_C$ for a single layer, as a function of
the density of holes,  for different values of the interdiffusion
parameter, $\Delta$, and for a fixed value  of the Mn concentration
$c_{M}=3.13\times 10^{14}$ cm$^{-2}$. We take the magnetization to be in
the plane of the layers in line with the experimental result.
\cite{Kawakami}	However, preliminary calculations show that the model
predicts an off-plane easy axis for the single digital layer, in the
idealized  $\Delta=0$ case. For each point we have calculated the
electronic structure  selfconsistently and solved Eq.~(\ref{integ})
to obtain $T_C$.  The general trend is that $T_C$ is an {\em
increasing} function of the  density of holes. The lines are the
best  fit  using  $T_c \propto p^{\beta}$.  In the inset we plot
$\beta$ as a function of $\Delta$.  In a  two-dimensional system with
parabolic bands, no spin-orbit interaction, and no Coulomb
interaction, we would have obtained $\beta=0$. Remarkably, $\beta$
in  the single layer is even larger than the value for bulk,
$\beta=1/3$. \cite{Dietl00}

In Fig.~\ref{fig3} we plot $T_C$ for the same set of single layers,
as a function of $\Delta$, for different densities of holes. The
model shows that interdiffusion reduces $T_C$.  For a fixed value of
$J$, there is a line in the  plane $(p,\Delta)$ which gives the same
$T_C$. In the inset of Fig.~\ref{fig3} we plot that line for both
$T_C=35$~K and $T_C=60$~K for  $J=150$~meV-nm$^3$. The first (35K)
corresponds to the Curie temperature reported by Kawakami {\em et
al.} \cite{Kawakami} to yield an idea of what model parameters could 
describe the experimental conditions.   The second (60K) has been
obtained for the same kind of heterostructures grown at slightly
higher temperature. \cite{private}

\section{Double digital layer}

In this section we report on our calculations of the electronic
structure and $T_C$ for two identical digital layers, separated by N
monolayers of GaAs, so that the interlayer distance is $d=N \times
0.2825$~nm.  Both layers are described by Eq.~(\ref{profile}). We
choose a point in the $(\Delta,p)$ parameter  space, so that, for
very large $d$, the calculated $T_C$ is close to  the experimental
value of 35 K.  Then we calculate $T_C(d)$ for smaller  values of
$d$.
\begin{figure}
\psfig{figure=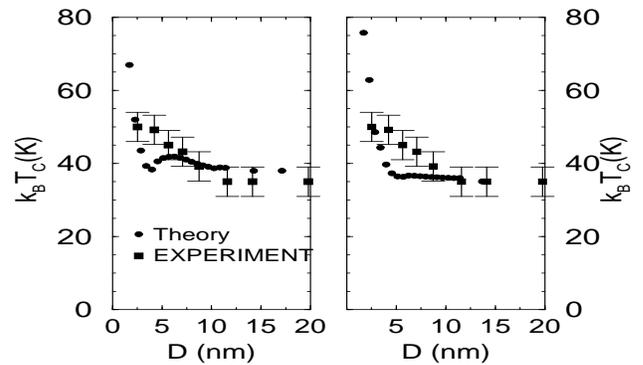,height=2in,width=3.3in}
\caption{Round dots: Curie Temperature as a function of the
interlayer distance for a double layer system. In the left panel,
$\Delta=0.5$~nm~ and $p=1.3\times 10^{13}$~cm$^{-2}$ . In the right
$\Delta=1.5$~nm~ and $p=3\times 10^{13}$~cm$^{-2}$. Square dots:
experimental Curie temperature for a multilayer
case.\protect{\cite{Kawakami}} }
\label{fig4}
\end{figure}

\begin{figure}
\psfig{figure=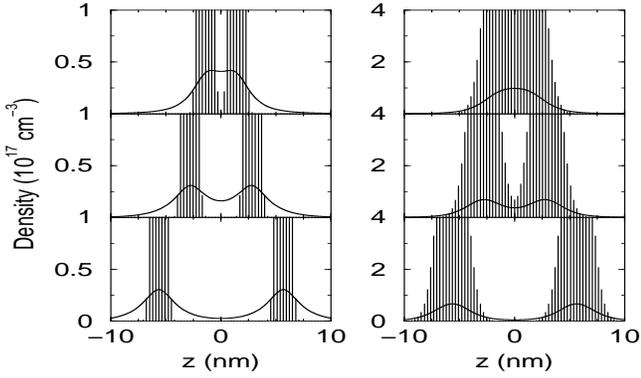,height=2in,width=3.3in}
\caption{Mn (shaded regions) and hole density (lines) profiles for a
double layer. Left panel:  $p=1.3\times 10^{13}$~cm$^{-2}$ per layer 
and $\Delta=0.5$~nm. Right panel:  $p=3\times 10^{13}$~cm$^{-2}$ per layer 
and $\Delta=1.5$~nm. From top to bottom, the interlayer distance is 10,
20 and 40 monolayers.}
\label{fig5}
\end{figure}

The results are shown in Fig.~\ref{fig4} for two cases: $(\Delta=
0.5$~nm, $p=1.3\times 10^{13}$~cm$^{-2})$ (left panel) and
$(\Delta=1.5$~nm, $p=3\times 10^{13}$~cm$^{-2})$ (right panel). The
theoretical results obtained with the first case give a better fit to
the experimental data \cite{Kawakami} than those obtained with the
second. In Fig.~\ref{fig5} we plot the corresponding density profiles
for 3 different interlayer distances to represent 3 regions of
separation dependence in the Curie temperature. At short layer
separations (upper  panels of Fig.~\ref{fig5}), both layers of the
hole and of the Mn distribution overlap and $T_c$ depends strongly on
the separation $d$. At  intermediate separations  (left middle panel
of Fig.~\ref{fig5}), the two layers of Mn  do not overlap but the
hole distribution still does. As a result, the layers  are coupled
and $T_C$ is weakly dependent on the interlayer distance. A  further
increase of $d$ leads to the uncoupled regime where $T_C$ reaches
the  single layer value (lower panels of Fig.~\ref{fig5}).

Both the calculated and the measured $T_C$  decrease as $d$ increases
and they reach a stationary value at large $d$. Similar  behavior is
obtained for several values of $(p, \Delta)$. We have also calculated
$T_C$ for 3, 4, and 5 delta layers and the  separation dependence is
similar. The steep decline of $T_C$ with separation stops  at
interlayer distances higher than about 10 monolayers, the relevant
experimental region. In all these cases, the theory seems to
slightly  {\em underestimate} the coupling between the layers at
intermediate  distances.  This might indicate that the density of
itinerant carriers in between  the magnetic layers might be higher in
the experiment than in our  calculations.  Further theoretical and
experimental work might clarify this point.

\section{ Digital layer inside a quantum well}

In this section we study the electronic structure and the $T_C$ of a
single digital ferromagnetic layer inside a quantum well, using the
formalism of sections II and III.  The model  predicts that the
confinement effect increases $T_C$ up to  a factor of 3 compared
with  the unconfined single layer. The system  consists of
Ga$_{x}$AsAl$_{1-x}$ barriers containing a GaAs well with a single
digital layer of Ga$_{0.5}$AsMn$_{0.5}$ in the middle. This structure
would be the ferromagnetic analog of the Be $\delta$-doped
GaAs/GaAsAl quantum well. \cite{BeQW}

\begin{figure}
\centerline{\psfig{figure=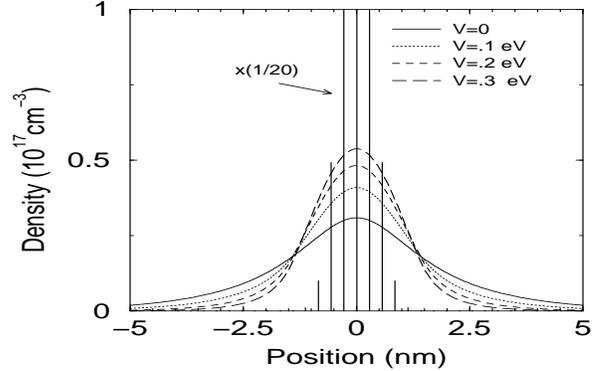,height=2in,width=3.0in}}
\caption{Hole density profiles for  different values of the barrier
potential, $V$, for a Mn layer (vertical lines) inside a quantum
well. The total density of holes is $p=1.3 \times10^{13}$ cm$^{-2}$. For the
sake of clarity, the Mn distribution has been divided by 20.}
\label{fig6}
\end{figure}

\begin{figure}
\centerline{\psfig{figure=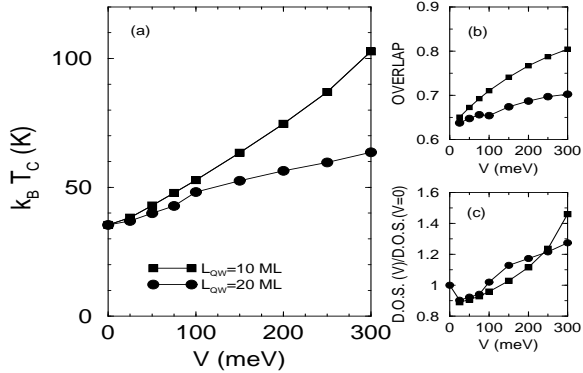,height=2in,width=3.0in}}
\caption{(a) Curie Temperature of a single digital layer in a quantum
well as a function of the barrier height, $V$, for two different values of
the well width, $L_{QW}$. (b) Overlap of the hole and the Mn
distributions as a function of  $V$. (c) Density of States (DOS) at the
Fermi level (in units of the DOS for $V=0$)}
\label{fig7}
\end{figure}

We model the interface GaAs/Ga$_{x}$AsAl$_{1-x}$ as a  barrier
potential of height $V= 550 \cdot x$ meV. The distribution of Mn is
that of  section IV, with $\Delta=0.5$nm, $p=1.3\times 10^{13}$~cm$^{-2}$.
In Fig.~\ref{fig6} we show the distribution of holes of a 10
monolayers wide quantum well for different values of the aluminum
content. As the barrier potential increases the hole distribution
overlaps more the Mn layer, increasing the $T_C$ . In Fig.~\ref{fig7}
we plot $T_C$ as a function of the barrier height (proportional to
the  Al content of the barrier), for a quantum well width of both 10
and 20 monolayers (2.8 and 5.7~nm)  with a  Mn layer in the middle
of the well,  with ($\Delta=0.5$ nm, $p=1.3\times
10^{13}$~cm$^{-2}$).  For the narrower well the enhancement factor
can be as large as 2.9 for $V=300$ meV which corresponds to 54\% of
Aluminum in the barriers. For a wider quantum well the effect is
smaller.

The increase of $T_C$ is due to two factors. The first is the
increase in the overlap between the Mn and the hole distributions
(see Fig.~\ref{fig6}). In Fig.~\ref{fig7}b we plot the overlap of the
Mn and the hole distributions, $\int dz \sqrt{ p(z) c_{M}(z)} /\sqrt{
p c_M}$, as a function of the barrier height, $V$. The increase of
the overlap due to confinement is larger for the narrower well. The
second factor is the increase of the density of states at the Fermi
Level (DOS).   In figure ~\ref{fig7}c we plot the DOS as a function
of $V$, normalized by the DOS at the Fermi level for the case $V=0$. At $V=0$ 
the Fermi Level is close to the bottom of the second  heavy hole band
so that there  3 bands are occupied (see figure 1) . The effect of
the barrier potential is to increase the energy level spacing so that
as $V$ increases, the Fermi level goes below the second Heavy hole
band and moves towards  the bottom of the first light hole  band. For
small values of $V$ the DOS at the Fermi level decreases slightly,
increasing up to 40$\%$ for higher values of $V$.

For $V>350$ meV and $L_{QW}=$10 nm, the Fermi Level gets close
to the bottom of the light hole band where there is a dramatic
increase of the DOS at the Fermi Level. This leads to an even larger
increase of the  predicted Critical Temperature. Work is in progress
to check if this result (not shown in the figures) remains  when the
finite  temperature of the fermions is taken into account.

 We have also checked that, for a 10 monolayer quantum well with a
density of holes of $p=2.5\times 10^{13} $ cm$^{-2}$, the relative increase
of  $T_C$ for $V$=300 meV, is a factor of $1.8$, {\em i.e.}, smaller
than the enhancement for $p=1.3\times 10^{13}$ cm$^{-2}$

\section{ Discussion and Conclusions}

The work presented in this paper is based on a model of
effective-mass, virtual-crystal and mean-field approximations. An 
important feature of the theory is that $T_C$ scales with the square
of the  exchange coupling constant, $J$. \cite{Dietl97,Dietl00,Mac00}
We have used a  value of $J=150$~meV-nm$^3$ from
Ref.~\onlinecite{Ohno98b,Mac00}. However, a  value 3 times smaller
has also been reported. \cite{newexp} This would change  $T_C$ by a
factor of nine. In turn, this could be compensated in the theory  by
increasing the density of holes, including the split-off band in the
calculation, \cite{Dietl00} a larger effective mass associated with 
the motion in an impurity band, \cite{Bhat} or a direct
ferromagnetic  coupling $J_M$ between the Mn atoms.

These caveats indicate the current difficulties with the theoretical
prediction of the absolute values of the Curie temperature.  Our
model  is less ambitious and is used to provide {\em changes} in the
behavior of  the Curie temperature as a function of parameters which
characterize the planar heterostructure. For instance, we have
presented an analysis of  $T_C$ for a single digital layer as a
function of the density of compensating impurities and the spread of
the Mn atoms due to interdiffusion. Our  results indicate that $T_C$
increases with the density of holes and with less interdiffusion
(smaller $\Delta$). The fact that $T_C$ depends on  density in a
quasi-two dimensional system, contrary to a naive calculation with
parabolic bands, gives some theoretical support to the experiment by
H.  Ohno {\em et al.} \cite{Naturedec2000} in which they observed a
change of  $T_C$ as the density of holes changes in a InAsMn quantum
well in  a  field-effect transistor. Our theory could be also applied
to model ferromagnetism observed in a quantum well of p-doped CdTeMn.
\cite{Haury97}

In section V we have presented our calculations for the double layer
system together with the experimental results for multilayers. The
qualitative agreement is good, but the theory seems to underestimate
the $T_C$ at intermediate interlayer distances, i.e., the coupling
between the magnetic layers is larger in the experiment than in the
theory. Further work on this problem, both in experimental
characterization and in the improvement  of the theory, might shed
some light on the microscopic origin of  ferromagnetism in this kind
of systems.

In section VI we have studied the confinement effects on the hole 
carriers which mediate the Mn-Mn magnetic coupling. This leads to a
prediction  of an increase of $T_C$ by as much as a factor of almost
3, when a single digital  layer is grown in a quantum well structure.
The capability to investigate the systemic changes on ferromagnetism
and to predict observable effects is  a strong point of the
effective-mass mean-field theory.

In conclusion, we have presented a theoretical framework to calculate
the electronic structure and the critical temperature of
heterostructures of III-V ferromagnetic semiconductors. This work is
an  extension of the 3D case, \cite{Dietl00} in which the relevance
of the spin-orbit interaction has been pointed out. The  main
features of the formalism, absent in  previous papers on 
heterostructures, are the inclusion of several subbands, necessary
because of the high density of holes in the system, and the inclusion
of the spin-orbit interaction, important because it changes both the
magnetic coupling  and the shape of the bands.  We have presented
calculations of  the digital magnetic heterostructures, providing a
qualitative  understanding of the experimental values of $T_C$, and
we have predicted that $T_C$ for a single digital layer can increase
by a factor of 2 when embedded in a quantum well.

We wish to thank Drs.~D.D. Awschalom, R. Kawakami, and A. Gossard for
stimulating discussions and Dr E. Gwinn for suggesting the calculation
of section VII. We acknowledge Spanish Ministry of Education
for a post-doctoral fellowship and support by DARPA/ONR
N0014-99-1-109  and NSF DMR 0099572.



\begin{references}

\bibitem{Spintronics} G. Prinz, Physics Today {\bf 45}, 58 (1995);
G.A. Prinz, Science {\bf 282}, 1660 (1998).

\bibitem{Ohno96} H. Ohno, {\em et al.}, Appl. Phys. Lett. {\bf 69},
363 (1996).

\bibitem{Esch97} A. Van Esch {\em et al.},  Phys. Rev. B {\bf56},
13103 (1997).

\bibitem{Ohno98a} H. Ohno, Science {\bf 281}, 951 (1998).

\bibitem{Ohno98b} F. Matsukura {\em et al.}, Phys. Rev. B {\bf57},
R2037 (1998).

\bibitem{Ohno00} H. Ohno, J. Magn. Magn. Mater. {\bf 200}, 110 (1999).

\bibitem{Furd} J.K. Furdyna, J. Appl. Phys. {\bf 64}, R29 (1988).


\bibitem{spinjec1} R. Fiederling {\em et al.}, Nature {\bf 402}, 787
(1999).

\bibitem{spinjec2} A.T. Filip {\em et al.} Phys. Rev. B {\bf 62}, 9996
(2000).

\bibitem{spinjec3} V.P. LaBella {\em et al.}, Science {\bf 292}, 1518
(2001).

\bibitem{spinjec4} A. Hirohata {\em et al.} Phys. Rev. B {\bf 63},
104425 (2001).


\bibitem{spinjec5} Y. Ohno {\em et al.}, Nature {\bf 402}, 790 (1999).

\bibitem{spinjec6} B. T. Jonker {\em et al.} Phys. Rev. B {\bf 62},
8180 (2000).

\bibitem{Dietl97} T. Dietl. A. Haury and Y. Merle d'Aubign\'e,
Phys. Rev.  B {\bf55}, R3347 (1997).

\bibitem{Dietl00} T. Dietl {\em et al.}, Science {\bf 287}, 1019
(2000).  T. Dietl, H.Ohno and F. Matsukura
Phys.  Rev.  B{\bf 63} 195205 (2001)

\bibitem{koshi} S. Koshihara, et al., Phys. Rev. Lett. {\bf 78}, 4617
(1997).

\bibitem{Naturedec2000}   H. Ohno et al., Nature {\bf 408}, 944
(2000).

\bibitem{Akiba} N. Akiba {\em et al.},  Appl. Phys. Lett. {\bf 73},
2122 (1998).

\bibitem{Kawakami} R. K. Kawakami {\em et al.}, Appl. Phys.
Lett. {\bf 77},  2379 (2000).

\bibitem{Deltabook} {\em Delta-doping of semiconductors}, edited by
E.F. Schubert, (Cambridge University Press, Cambridge, 1996).

\bibitem{private} R. K. Kawakami and A. Gossard, private communication.

\bibitem{KS} W. Kohn and L.J. Sham, Phys. Rev. {\bf 140}, A1333 (1965).

\bibitem{Broido} D. Broido and L.J. Sham, Phys. Rev. B {\bf 31},
888 (1985).
\bibitem{KohnLutt} J.M. Luttinger, Phys. Rev. {\bf 102}, 1030, (1956).

\bibitem{Mac00} T. Jungwirth {\em et al.}, Phys. Rev. B {\bf 59},
9818 (1999).

\bibitem{MacQW} B. Lee, T. Jungwirth and A.H. MacDonald, Phys.
Rev. B {\bf61}, 15606 (2000).

\bibitem{brey} L. Brey, F. Guinea, Phys. Rev. Lett {\bf 85}, 2384
(2000)

\bibitem{nosoqw} A. Ghazali, I.C. da Cunha Lima and M.A. Boselli,
Phys. Rev. B {\bf63}, 153305-1 (2001).


\bibitem{Mac00b} J. K\"onig, H. Lin and A.H. MacDonald, Phys. Rev.
Lett. {\bf 84}, 5628 (2000). 

\bibitem{Yangcomm} M. Yang, S. Sun and M. Chang, Phys. Rev.
Lett. {\bf86}, 5636 (2001) 

\bibitem{KL} M. Sigrist, K. Ueda and H. Tsunetsugu, Phys. Rev.
B {\bf46}, 175 (1992)

\bibitem{yang} S.-R. Yang, D. Broido, and L.J. Sham, Phys. Rev. B {\bf 
32}, 6630 (1985).

\bibitem{Lazzouni} M.E. Lazzouni and L. J. Sham,  The
International Journal for Computation and Mathematics in
Electrical and Electronic Engineering, {\bf 14}, 129 (1995).

\bibitem{BeQW} Y.C. Shih and B.G. Streetman, Appl. Phys. Lett. {\bf
59}, 1344 (1991). J. Wagner and D. Richards , Chapter 15 in
Ref.~\onlinecite{Deltabook}.

\bibitem{newexp} T. Omiya {\em et al.}, Physica E{\bf 7}, 976
(2000).

\bibitem{Bhat}  R.N. Bhatt and M. Berciu, cond-mat/0011319.

\bibitem{Haury97} A. Haury {\em et al.}, Phys. Rev. Lett. {\bf 79},
511 (1997).

\end{references}
\end{document}